\documentclass[a4paper,11pt]{article}
\usepackage{graphicx}
\usepackage{epstopdf}
\usepackage{amsfonts,amsmath,amssymb}
\usepackage{hyperref}
\usepackage{float}

\usepackage{epsfig,multicol,bbm}

\newcommand\fverb{\setbox\pippobox=\hbox\bgroup\verb}
\newcommand\fverbit{\egroup\item[\fbox{\unhbox\pippobox}]}

\newbox\pippobox

\begin{document}
\title{\bf  Kaluza-Klein   Magnetized Cylindrical Wormhole and its Gravitational Lensing }
\author{ S. Sedigheh Hashemi\,\, and\,\, Nematollah Riazi\thanks{Electronic address: n\_riazi@sbu.ac.ir}
\\
\small Department of Physics, Shahid Beheshti University,  Evin, Tehran 19839,  Iran}
\maketitle
\begin{abstract}
A new exact vacuum solution in five dimensions, which describes a magnetized cylindrical wormhole in $3+1$ dimensions is presented. The magnetic field lines are stretched along the wormhole throat and are concentrated near to it.
 We  study the motion of neutral and charged test particles under the influence of the magnetized wormhole. The effective potential for  a neutral test particle around and across the magnetized wormhole   has a repulsive character. The gravitational lensing for the magnetized wormhole for various lens parameters are calculated and compared. The total magnetic flux on either side of the wormhole is obtained.
  We present analytic expressions which show regions in which the null energy condition is violated.
\end{abstract}
\maketitle
\section{Introduction}\label{1}

The first brilliant insight for unifying the Einstein\rq{}s theory of gravitation and Maxwell\rq{}s theory of electromagnetism
was proposed by Kaluza and Klein  by considering a pure gravitational theory in a five-dimensional  manifold \cite{0}.
 The Kaluza-Klein theory of gravity and electromagnetism has been the starting point for the development of higher dimensional theories of gravity including the superstring and the M-Theory \cite{C. M. Chen}.

Most of the exact solutions of Kaluza-Klein theory including black holes, electric pp-waves, \cite{-1}  magnetic monopoles  and  solitons have been studied (See \cite{-2},\cite{riazi} and references therein).

Wormholes are strongly gravitating objects connecting remote parts of the universe or one distant region of spacetime to another\cite{1}, \cite{-3}.
 They are studied for several purposes such as describing physics as pure geometry and as a model for a charged particle represented by a bridge  connecting two asymptotically Minkowskian manifolds called Einstein-Rosen bridge \cite{2}.
 The general theory of a traversable Lorentzian wormhole was formulated by Morris and Thorne \cite{3}. Wormholes suggested by Morris and Thorne are spherically symmetric with a smooth surface throat \cite{-10}. They also showed that this kind of wormhole not only might allow humans to travel among universes, but also to build time machines \cite{-9}.
 Another application of wormholes is in the development of quantum gravity \cite{4}.
Static  Lorentzian wormholes in four-dimensional Einstein gravity can exist only in the presence of  exotic matter which violates the null energy conditions \cite{5},\cite{-111}. This is a challenging feature in wormhole physics, which is named  as the \lq\lq{}exotic matter problem\rq\rq{}. Attempts for solving this issue includes alternative theories of gravity \cite{-12}-\cite{-14}, using quantum effects in curved spacetimes and regarding wormholes as semi classical objects \cite{-15}. Another approach for solving this problem  is to consider wormholes in higher dimensional theories of gravity, in which  exotic matter is not obligatory \cite{6}.

The first Kaluza-Klein wormhole was discovered by Chodos and Detweiler \cite{-4}. The authors obtained a class of spherically symmetric and asymptotically flat solutions describing  wormholes. Moreover, this category of solution was extended to axisymmetric multi-wormholes \cite{-5}. It is shown in refs. \cite{00} and \cite{01}  that by parametrizing the off-diagonal metrics in five dimensional Kaluza-klien theory the new vacuum Einstein solutions will be defined. These solutions are wormholes which are locally anisotropic. Cylindrically symmetric Abelian wormholes in $(4+n)$-dimensional Kaluza-Klein theory is obtained in \cite{02}.

 In this article, we present a vacuum  solution within the framework of five-dimensional Kaluza-Klein gravity  which reduces to a magnetized cylindrical wormhole  in four-dimensional curved spacetime.
The main motivation behind the present contribution is to explore the physical properties and manifestations of magnetized cylindrical wormholes. In fact, we consider a wormhole which is topologically   different from the Morris and Thorne wormhole, starting with the cylindrical symmetry assumption.

 Our  paper is laid out  as follows: In the next section, we present a five-dimensional metric which satisfies the vacuum Einstein equations in 4+1 dimensions. The standard (3+1)D reduction is then carried out to show that the solution describes a magnetized cylindrical wormhole.
After this reduction, the circumference of the magnetized wormhole is obtained and  the radius of the throat is calculated.
It is instructive to study the motion of a test particle in the electromagnetic field of the  magnetized  wormhole by using the  Lagrangian equations which are presented in section (\ref{3}). We derive the effective potential  for the motion of a test particle around the magnetized wormhole which has a repulsive behavior.
We also present the radial motion and the equation of orbit for some special cases.  The same procedure is done for a charged particle. In section (\ref{4}), we discuss gravitational lensing of the magnetized wormhole in the weak field limit.
We calculate the magnetic flux of the wormhole in section (\ref{5}). The energy conditions are  analysied in section (\ref{6}). Finally in the last section we will draw our conclusions.
\section{Magnetized Cylindrical Wormhole Solution}\label{2}
Let us start with a five-dimensional metric in  the form
\begin{equation}\label{1}
 {\rm d}s^2= -f(r) {\rm d}t^2+ {\rm d}r^2+ {\rm d}z^2-h(r) {\rm d}w^2+p(r) {\rm d}w {\rm d}\theta,
\end{equation}
where the  extra coordinate is $w$ and the  three dimensional cylindrical coordinates are given by $r$, $z$, $\theta$ with usual ranges. We require that all functions $f$, $h$ and $p$ be functions of $r$. Clearly $\partial _{t}$, $\partial _{z}$, $\partial _{\theta}$ and $\partial _{w}$ are the killing vectors of the given metric. It can be verified from the suggested metric (\ref{1}) that by having $t, r$ and $z={\rm const}$ hypersurfaces, a 2-dimensional sub space will be obtained. In this sub manifold by setting $w={\rm const}$ the line element ${\rm d}s^2$ will be zero. Therefore, all $w={\rm const}$ trajectories are null. We will show further that the cross term $ {\rm d}w {\rm d}\theta$ in the metric is a necessity for having a magnetic field in the four dimensional space time. We are interested  in vacuum solutions in five dimensions $(R_{AB}=0)$. The vanishing Ricci tensor leads to the following set of coupled, non-linear differential equations
\begin{equation}\label{eq2}
R_{tt}=\frac{1}{2}f^{\prime \prime  }-\frac{1}{4}\frac{ f ^{\prime  2} }{f}+\frac{1}{2}\frac{    f ^{\prime  }  p ^{\prime  }   }{p}=0,
\end{equation}
\begin{equation}
R_{rr}=-\frac{1}{4}\left(\frac{f ^{\prime}   }{f}\right)^2 -\frac{1}{2}\frac{f^{\prime \prime  } }{f}+\frac{1}{2}\left(\frac{p ^{\prime} }{p}\right)^2-\frac{p^{\prime \prime 2} }{p}=0,
\end{equation}
\begin{equation}
R_{ww}=-\frac{1}{4}\frac{      f ^{\prime}h^{\prime}            }{f}-\frac{1}{2}h^{\prime \prime}-\frac{1}{2}\left(\frac{p ^{\prime} }{p}\right)^2 h+\frac{1}{2}\frac{p^{\prime}  h^{\prime}  }{p}=0,
\end{equation}
\begin{equation}
R_{\theta w}=-\frac{1}{8}\frac{f ^{\prime}  p^{\prime}   }{f}-\frac{1}{4}p^{\prime \prime}=0.
\end{equation}
In order to have a zero Ricci tensor it is necessary (but not sufficient) to have vanishing Ricci scalar given by
\begin{equation}\label{eq6}
R=-\frac{f^{\prime \prime  } }{f}+\frac{1}{2}\left(\frac{f ^{\prime}   }{f}\right)^2-\frac{f ^{\prime}  p^{\prime}   }{fp}+\frac{1}{2}\left(\frac{p ^{\prime} }{p}\right)^2-\frac{2p^{\prime \prime} }{p}=0.
\end{equation}
It is seen that the Ricci scalar is independent of the function $h(r)$, that is $\frac{\partial R}{\partial h}=0$.
Solving the set of Eqs.\,(\ref{eq2})-(\ref{eq6}) leads to the metric
 \begin{align}\label{eq7}
 {\rm d}s^2 _{(5)}=&-\frac{c}{\left(ar\right)^\frac{2}{3}}{\rm d}t^2  +{\rm d}r^2+{\rm d}z^2-     r^\frac{4}{3}\left(c+d\ln{r}\right){\rm d}w^2   +2\left(ar\right)^\frac{4}{3}{\rm d}w{\rm d}\theta,
 \end{align}
where  $c, a$ and $d$ are constants of integration. Here, the  $t$ and $r$ constant hypersurfaces are assumed to have $T^2\times R$ topology.

The theory of Kaluza-Klein compactification  in five dimensions is to unifying gauge fields and gravity by compactifying the extra dimension in a higher dimensional pure gravity \cite{12}. One can decompose the five-dimensional metric in the string-frame as
  \begin{equation}\label{eq8}
(\hat g_{AB})= \left(
‎\begin{array}{ccccccc}‎
‎g_{\mu \nu}+\kappa^2 \phi^2 A_{\mu} A_{\nu}‎   & \kappa \phi^2 A_{\mu} &  \\‎
 ‎‎  &‎\\‎
\kappa \phi^2 A_{\nu}    ~~& \phi^2 &  \\‎
‎\end{array}
‎\right),
\end{equation}
here $g_{\mu \nu}$ is the four-dimensional metric tensor, $\phi$ is the scalar field, $A_{\mu}$ is the electromagnetic potential and $\kappa$ is the coupling constant, the Latin  indices $A,B$ run over $0...4$ and the Greek indices $\mu, \nu$ run over $0...3$.
It is assumed that the extra dimension is curled up into a very small circle of radius $R  ~ (x^5=x^5+2\pi R, M^5\rightarrow M^4 \times S^1)$ which is not observable \cite{13}.

In the tarditional Kaluza-Klein decomposition of the five-dimensional metric,  a  scalar field which comes from the
 $g_{44}$ component of the five-dimensional metric, a four-dimensional spacetime plus an  electromagnetic
 field which is a source term for the scalar field can be identified from  (\ref{eq8}).
Consequently,
the  scalar field and the electromagnetic field resulting from metric (\ref{eq7}) are  given by
  \begin{equation}
  \phi^2=|r^\frac{4}{3}\left(c+d\ln r \right)|,
  \end{equation}
  and
  \begin{equation}
  A_{\theta}=\frac{a^\frac{4}{3}}{\kappa |c+d\ln {r}|},
  \end{equation}
  respectively. If $\phi=0$, the entire system reduces to the four-dimensional Einstein gravity without any electromagnetic field. Since without the scalar field $\phi$  exact solutions will not exist, people call it the dilaton \cite{03}. The only non-vanishing component of the electromagnetic field tensor $F_{\mu \nu}=\partial _{\mu}A_{\nu}-\partial _{\nu}A_{\mu}$ is
  \begin{equation}
  F_{r \theta}=-F_{\theta r}=-\frac{da^\frac{4}{3}}{\kappa r\left(c+d\ln r\right)^2},
  \end{equation}
  which corresponds to a magnetic field along the $z$ axis.
Moreover, the reduction from five-dimensional metric (\ref{eq7})  down to four-dimensional spacetime with the help  of Eq. (\ref{eq8})  leads to the following metric in cylindrical coordinates
\begin{equation}\label{eq12}
{\rm d}s^2_{(4)}=-\frac{c}{(ar)^\frac{2}{3}  }{\rm d}t^2+ {\rm d}r^2+
\frac{a^\frac{8}{3} r^\frac{4}{3}}{|c+d\ln{r}|}{\rm d}\theta^2+{\rm d}z^2.
\end{equation}
By introducing a new time parameter $ \tilde{t}=at $, Eq. (\ref{eq12})   can be  rewritten as  
\begin{equation}
{\rm d}s^2_{(4)}=-\frac{1}{\tilde{a}r^{\frac{2}{3}}}{\rm d}\tilde{t}^2+{\rm d}r^2+\frac{\tilde{a}r^{\frac{4}{3}}}{|1+\tilde{d}\ln{r}|}{\rm d}\theta^2+{\rm d}z^2,
\end{equation}
where $\tilde{a}\equiv \frac{a^{\frac{8}{3}}}{c}$ and   $\tilde{d}=\frac{d}{c}$.

Let us consider  $\tilde{t}$ and $z$ constant hypersurfaces and calculate $C(r)$ which is the circumference  of a circle with $r=\textrm{constant}$,  that is
\begin{equation}
C(r)=\int ^{2\pi }_{0} \frac{\sqrt{ \tilde{a}}r^\frac{2}{3}}{\sqrt{|1+\tilde{d}\ln{r}|}}{\rm d}\theta=\frac{2\pi \sqrt{ \tilde{a}}r^\frac{2}{3}}{   \sqrt{|1+\tilde{d}\ln{r}|      }        }
\end{equation}
We  present in Figure. \ref{Cir2}, the behavior of the function $C(r)$ for three values of the parameters $a, c$ and $d$.
As it is seen in the figure, the circumference $C(r)$ has a  minimum which can be obtained  via
\begin{equation}
\frac{dC(r)}{dr}=0,
\end{equation}
 which gives
\begin{equation}
r_{th}\equiv r_{min}=e^ {      \frac{3}{4}-\frac{1}{\tilde{d}}},
\end{equation}
identified as the radius of the magnetized cylindrical wormhole\rq{}s throat.  One can also see that the minimum circumference condition comes from the minimum area condition for the throat \cite{-17} and the fact that the throat has cylindrical symmetry.

The so called \lq\lq{}flare out\rq\rq{} condition which is a key requirement at the throat must be satisfied. The generalization of the Morris-Thorne flare out condition for a static wormhole is given by  \cite{-17}
\begin{equation}\label{100}
\frac{\partial {\rm tr(K)}}{\partial n}\leq 0,
\end{equation}
where $n$ is the Gaussian normal coordinate $x^i=(x^a;n)$  and $K$ is the trace of the extrinsic curvature in Gaussian normal coordinates which is defined by
\begin{equation}
K_{ab}=-\frac{1}{2}\frac{\partial g_{ab}}{\partial n}.
\end{equation}
For computing the above equations, we define a two-dimensional hypersurface $\Sigma$ which has a minimum area taken in constant time slices. In the  Gaussian normal coordinates, $\Sigma$ lies at $n=r-r_{th}$, therefore we can decompose the line element as
\begin{equation}
{\rm d}s^2_{(3)}={\rm d}s^2_{(2)}+{\rm d}n^2,
\end{equation}
where
\begin{equation}
{\rm d}s^2_{(2)}=\frac{\tilde{a}r^{\frac{4}{3}}}{|1+\tilde{d}\ln{r}|}{\rm d}\theta^2+{\rm d}z^2.
\end{equation}
Now Eq. ({\ref{100}) gives
\begin{equation}
\frac{\partial {\rm tr(K)}}{\partial n}=-\frac{\tilde{a}}{2}\frac{\partial^2}{\partial r^2}(\frac{r^{4/3}}{|1+\tilde{d}\ln{r}|})\mid_{r=r_{th}},
\end{equation}
which gives a  negative value after inserting $r_{th}$ in the above equation. So that the flare out condition is satisfied by our wormhole solution.

\begin{center}
\begin{figure}[H] \hspace{4.cm}\includegraphics[width=8.cm]{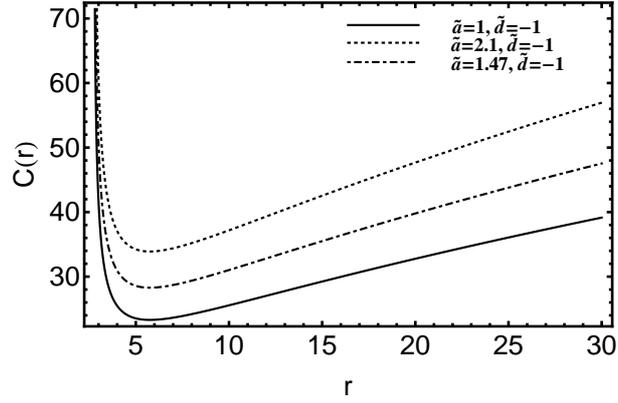}\caption{\label{Cir2} \small
The plot depicts the circumference of a  circle with $r=\textrm{constant}$ for different values of the parameters $\tilde{a}$ and $\tilde{d}$. The  minimum is identified as the radius of the wormhole\rq{}s throat.}
\end{figure}
\end{center}
\begin{center}
\begin{figure}[H] \hspace{4.cm}\includegraphics[width=8.cm]{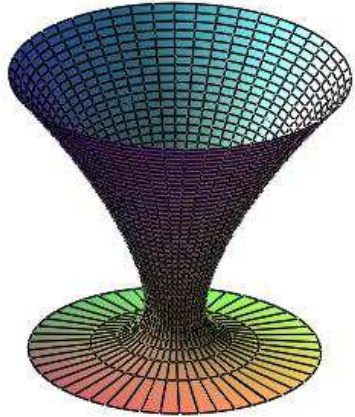}\caption{\label{wh} \small
Visualization of the magnetic wormhole. The plot is obtained by the requirement that the circumference of horizontal circles equals $C(r)$ and the vertical axis is $r$.}
\end{figure}
\end{center}
\section{Geodesic Motion of a Neutral Test Particle }\label{3}
In this section, we begin by obtaining the geodesic motion of a test particle near the magnetized cylindrical wormhole which can  be derived from Lagrangian in $(3+1)$ dimensions
  \begin{equation}\label{16}
   \mathfrak{L}=\frac{1}{2}g_{\mu \nu}\dot x ^\mu \dot x ^ \nu=\frac{1}{2}\{ -\frac{1}{\tilde{a}r^{\frac{2}{3}}}\dot{\tilde{t}}^2+\dot{r}^2+\frac{\tilde{a}r^{\frac{4}{3}}}{|1+\tilde{d}\ln{r}|}\dot{\theta}^2+\dot{z}^2 \}.
  \end{equation}
Throughout  this paper  an overdot denotes differentiation with  respect to an affine parameter that could be the proper time $(\tau)$ for time-like geodesics and affine parameter $(\lambda)$ for the null case. Since the Lagrangian is static and does not depend explicitly on the  coordinate $z, \tilde{t}$ and the azimuthal coordinate $\theta$, these symmetries give rise to three integrals of motion
given by
\begin{equation}\label{eq17}
-\frac{\tilde{a}}{r^\frac{2}{3}} \dot{\tilde{t}}=E,
\end{equation}
\begin{equation} \label{18}
\frac{\tilde{a}r^{\frac{4}{3}}}{|1+\tilde{d}\ln{r}|}\dot{\theta}=L.
\end{equation}
and
\begin{equation}
 \dot{z}= P=constant,
\end{equation}
where the constants of integration $E$, $L$ and $P$ can be regarded as the energy,  angular momentum and  linear momentum per unit mass in the $z$ direction, respectively.
The resulting  equation of motion corresponding to $r$ is
\begin{equation}\label{20}
 \ddot{r}-\frac{E^2}{3\tilde{a}^3 r^{\frac{1}{3}}}+\frac{L^2}{6\tilde{a}}\frac{(3\tilde{d}+4(1+\tilde{d}\ln{r}))}{r^{\frac{7}{3}}}=0,
 \end{equation}
 which can be integrated to give
 \begin{equation}
 \frac{1}{2}\dot{r}^2+V(r)=\epsilon
 \end{equation}
where the quantity  $V(r)$ can be thought as   the  effective  potential for the  motion of a test particle and $\epsilon$  is a constant of integration, regarded as the total energy of the test particle per unit mass. Combining these equations we can solve for the potential energy as a function of the radial coordinate $r$
\begin{equation}\label{eq22}
V\left(r\right)=\frac{3\tilde{a}\tilde{d}L^2+4E^2 r^2-\tilde{a}^2L^2(4+3\tilde{d}+4\tilde{d}\ln{r})}{8\tilde{a}^3 r^{\frac{4}{3}}}
\end{equation}
As the potential function involves unspecified parameters, it is difficult to proceed analytically. Figure \ref{pot4}
shows the radial dependence of the potential energy of the test particle  for different values of the parameters  $A$, $B$, $\tilde{a}$ and $\tilde{d}$, where for simplicity we define $A\equiv \frac {L^2}{8\tilde{a}^3}$ and $B \equiv \frac{E^2 }{2\tilde{a}^3}$. The  behavior of the orbits can be understood by comparing $\epsilon$ to $V(r)$.
The particle will move in the potential until it reaches the turning point where $V(r)=\epsilon$, and then it will begin moving in the other direction.
Since for each of the cases considered in   Figure \ref{pot4}, the maximum of the effective potential is located at $r_{max}<r_{th}$, therefore the potential has a repulsive character
 which  means that when a particle comes from infinity and passes by the magnetized wormhole throat it  will  be either reflected and go  to infinity again, or pass through the wormhole, depending on the value of the angular momentum $L$.

\begin{center}
\begin{figure}[H] \hspace{4.cm}\includegraphics[width=8.cm]{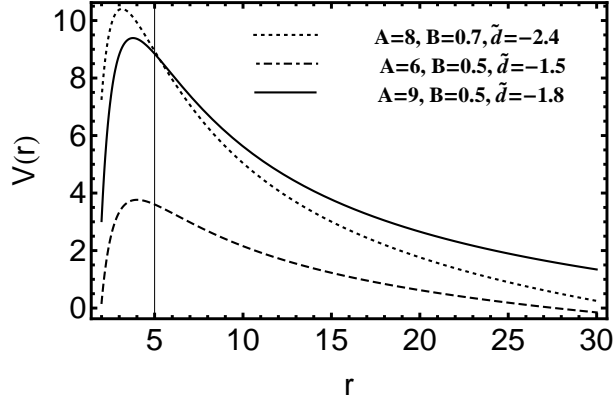}\caption{\label{pot4} \small
Radial dependence of the potential energy for different values of the parameters $A$, $B$, $c$ and $d$. Throat of the wormhole $r_{th}$ for these values is in the range $3<r_{th}<4.2$. The potential has a repulsive character.}
\end{figure}
\end{center}
Now we can specialize to the case of the motion in the equatorial plane and  radial geodesics defined by $z, \theta={\rm const}.$ From (\ref{16}) we get
\begin{equation}
2\mathfrak{L}=  -\frac{1}{\tilde{a}r^{\frac{2}{3}}}\dot{\tilde{t}}^2 +\dot{r}^2.
\end{equation}
Radial motion of the neutral test particle in the field of the magnetized wormhole can be described using the following equation derived from (\ref{20}), with $L$ set to zero:
\begin{equation}\label{24}
 \ddot{r}-\frac{E^2 }{3\tilde{a}^3r^{\frac{1}{3}}}=0,
 \end{equation}\label{25}
 which can be easily integrated to give
 \begin{equation}
\frac{1}{2}\dot{r}^2-Kr^\frac{2}{3}=\frac{1}{2}\dot{r}^2+V(r)=\epsilon
 \end{equation}
 where $K$ is defined as $K\equiv \frac{E^2}{2\tilde{a}^3} $, $\epsilon$ is a constant of integration corresponding to the total energy and $V(r)$  is the potential energy (per unit mass) of the test particle. Figure  \ref{pot3} illustrates  the potential
   of the neutral test particles around the magnetized cylindrical wormhole for different values of the parameter $K$.
As it is seen from  Figure  \ref{pot3}, the potential has a repulsive character which shows that the particle can either be bounced back to infinity, or pass through the wormhole, depending on its energy $\epsilon$.

\begin{center}
\begin{figure}[H] \hspace{4.cm}\includegraphics[width=8.cm]{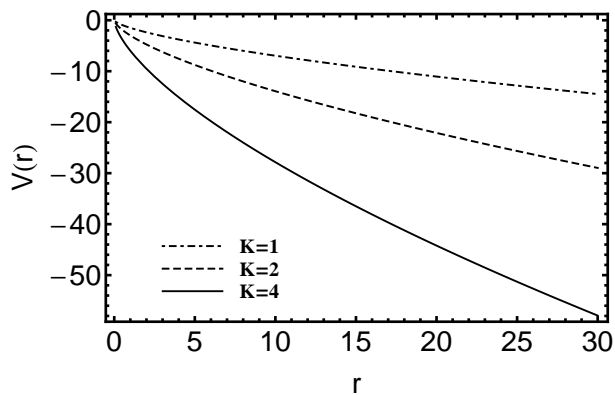}\caption{\label{pot3} \small
Radial dependence of the potential  for different values of the parameter $K$.  The potential has a repulsive character.}
\end{figure}
\end{center}
\subsection{Equation of Orbit and Constants of Motions}
In order to complete the discussion of the previous section, we calculate the equation of orbit for $r(\theta)$ in this section. Here we restrict our attention to the orbital motion of the test particle which moves in the $z={\rm const}$ plane. To get an equation for $r(\theta)$ we  start with
\begin{equation}\label{26}
r\rq{}(\theta)=\frac{{\rm d}r}{{\rm d}\theta}=\frac{\dot{r}}{\dot{\theta}},
\end{equation}
which  by using the constants of motion (\ref{eq17}) and (\ref{18})  leads to
\begin{equation}\label{27}
r\rq{}(\theta)^2=\dot{r}^2\frac{ \tilde{a}^2r^{\frac{8}{3}} }{       L^2     ( 1+\tilde{d} \ln{r} )^2 }.
\end{equation}
By dividing the metric (\ref{eq12}) by $({\rm d}\tau)^2$ and putting ${\rm d}z=0$, we obtain
\begin{equation}\label{28}
1=  -\frac{1}{\tilde{a}r^{\frac{2}{3}}}\dot{\tilde{t}}^2+\dot{r}^2+\frac{\tilde{a}r^{\frac{4}{3}}}{|1+\tilde{d}\ln{r}|}\dot{\theta}^2,
\end{equation}
Equations (\ref{27}) and (\ref{28}) then lead to
\begin{equation}\label{29}
r\rq{}(\theta)=\pm \{\tilde{a}\frac{r^\frac{4}{3}}{1+\tilde{d}\ln {r}} + \frac{E^2 }{L^2\tilde{a} }   \frac{r^\frac{10}{3}}{(1+\tilde{d}\ln {r})^2}   +\frac{\tilde{a}^2 }{L^2}    \frac{r^\frac{8}{3}}{(1+\tilde{d}\ln {r})^2}       \}^{\frac{1}{2}}.
\end{equation}
Since Eq. (\ref{29}) can not be integrated analytically
 to obtain the explicit form of the function $r(\theta)$, we have plotted
  $r\rq{}(\theta)$
 in  Figure \ref{EO}  for  $C=0.5$, $D=1$, $\tilde{d}=-11$ and $\tilde{a}=0.5$, where we have defined $C\equiv \frac{E^2 }{\tilde{a}L^2}$ and $D\equiv \frac{\tilde{a}^2 }{L^2}$. It can be seen in this figure that the function $r\rq{}(\theta)$ has a singularity at $r\approx 1$.
 \begin{center}
\begin{figure}[H] \hspace{4.cm}\includegraphics[width=8.cm]{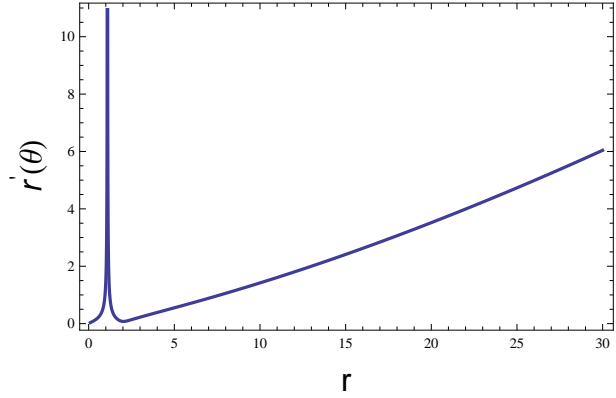}\caption{\label{EO} \small
The function $r\rq{}(\theta)$ for the test particle is plotted for the constants $C=0.5$, $D=1$, $\tilde{d}=-11$ and $\tilde{a}=0.5$, where we have defined $C\equiv \frac{E^2 }{\tilde{a}L^2}$ and $D\equiv \frac{\tilde{a}^2 }{L^2}$ in the Eq. (\ref{29}).}
\end{figure}
\end{center}
 Let us proceed with
 the motion of a charged particle with charge $e$ and mass $m$ in the gravitational and magnetic fields of the magnetized cylindrical wormhole.
The Lagrangian  is given by \cite{-11}
\begin{equation}\label{30}
  \mathfrak{L}=\frac{1}{2}g_{\mu \nu}\dot x^\mu \dot x^\nu-\frac{e}{m}A_{\mu}\dot x^\mu,
  \end{equation}
  since in our case the only component of the electromagnetic field is $A_{\theta}$, the Lagrangian  will reduce to
  \begin{equation}
   \mathfrak{L}=\frac{1}{2}\left[    -\frac{1}{\tilde{a}r^{\frac{2}{3}}}\dot{\tilde{t}}^2+\dot{r}^2+\frac{\tilde{a}r^{\frac{4}{3}}}{|1+\tilde{d}\ln{r}|}\dot{\theta}^2+\dot{z}^2    \right]
-\frac{e}{\kappa m}\frac{\sqrt{\tilde{a}}}{\sqrt{c} |1+\tilde{d}\ln{r}|}\dot{\theta}.
\end{equation}
Following the standard  procedure,  the constants of motion can be read as
\begin{equation}
\frac{\tilde{a}r^{\frac{4}{3}}}{|1+\tilde{d}\ln{r}|}\dot{\theta}-
 \frac{e}{\kappa m}\frac{\sqrt{\tilde{a}}}{\sqrt{c} |1+\tilde{d}\ln{r}|}=L,
\end{equation}
\begin{equation}
-\frac{\tilde{a}}{r^\frac{2}{3}} \dot{\tilde{t}}=E.
\end{equation}
Finally, by using the above constants of motion,
the equation of orbit $(r\rq{}(\theta))$  may be found which is a complicated function of the radial coordinate $r$ given by
\begin{align}
\ddot{r}&=\frac{E^2}{3 \tilde{a}^3 r^{\frac{1}{3}}}+\frac{\tilde{d}e}{\kappa m \sqrt{c\tilde{a}}r^{\frac{7}{3}}|1+\tilde{d}\ln{r}|}(L+\frac{e \tilde{a}}{m\kappa \sqrt{c}|1+\tilde{d}\ln{r}|})
\nonumber\\&-\frac{\tilde{d}}{2 \tilde{a}r^{\frac{7}{3}}}(L+\frac{e \tilde{a}}{m\kappa \sqrt{c}|1+\tilde{d}\ln{r}|})^2.
\end{align}
\section{Gravitational Lensing by the Wormhole}\label{4}
The first experimental confirmation of the Einstein theory of general relativity was the light bending \cite{120}.
At present, gravitational lensing plays a crucial role in astronomy and cosmology. By using the gravitational lensing, the cosmological constant, extrasolar planets, dark energy and dark matter can be probed (see Perlick \cite{121}, \cite{122} and Schreider et al. \cite{123}).

  The possibility of  astrophysical wormholes has been explored \cite{124}-\cite{126}. As an example, Kardashev et al. proposed that some active galactic nuclei may be explained as wormholes \cite{127}. Gravitational lensing effects by wormholes was pioneered by Cramer et al. \cite{128} and Kim and Cho \cite{129}. Since then, the gravitational lensing effects by different types of wormholes have been studied \cite{130}-\cite{137}.
  In this section, we calculate the lensing by the proposed magnetic wormholes.
  
To work out the lens equation we have to calculate the null geodesics in the   $z={\rm const}$ plane without loss of generality. The 
two constants of motion (\ref{eq17}) and (\ref{18}) are substituted into the null condition ${\rm d}s^2=0$ and using (\ref{27}) one can  obtain the following  equation for the photon orbit as
\begin{equation}\label{40}
r\rq{}(\theta)^2=\frac{E^2 }{\tilde{a}L^2}\frac{r^{\frac{10}{3}}}{(1+\tilde{d}\ln{r})^2}-\tilde{a}\frac{r^{\frac{4}{3}}}{|1+\tilde{d}\ln{r}|}.
\end{equation}
An integration of Eq. (\ref{40}) immediately gives the deflection angle expressed as 
\begin{equation}\label{42}
\theta=\pm \int\frac{{\rm d}u}{\sqrt{{ M \frac{u^{\frac{2}{3}}}{(\frac{3}{4}-\ln{u})^2}}  +N\frac{     u^{\frac{8}{3} } }{ (\frac{3}{4}-\ln{u}    )      }}}.
\end{equation}
We now make a coordinate transformation from $r$ $\in[r_{th}, +\infty)$ to $u \in [0,u_{m}) $ by $u=\frac{r_{th}}{r}$ and $M\equiv\frac{E^2}{\tilde{a}\tilde{d}^2L^2}r_{th}^{\frac{4}{3}}$, $N\equiv \frac{\tilde{a}}{\tilde{d}r_{th}^{\frac{2}{3}}}$.

The minimum distance $r_m$ that a light trajectory  has from the wormhole is obtained via $r\rq{}(\theta)=0$, which gives the following constraint 
\begin{equation}\label{43}
\frac{E^2}{\tilde{a}^2L^2}=\frac{|1+\tilde{d}\ln{r_m}|}{r_{m}^2}.
\end{equation}
By plugging Eq. (\ref{43}) in Eq. (\ref{42}), the deflection angle is rearranged as 
\begin{equation}
\theta=\pm \frac{r_{th}^{\frac{1}{3}}\sqrt{\tilde{d}}}{\sqrt{\tilde{a}}} \int^{\infty} _{u_{m}} \frac{{\rm d}u}{\sqrt {  \frac{r_{th}^2}{\tilde{d}}\frac{|1+\tilde{d}\ln{r_m}|} {   r_{m}^2              }    \frac{u^{\frac{2}{3}}}{(\frac{3}{4}-\ln{u})^2}                                   +\frac{u^{\frac{8}{3}}}{(\frac{3}{4}-\ln{u}    )}       }}.
\end{equation}
This is rewritten as
\begin{equation}\label{45}
\theta=\pm \frac{r_{th}^{\frac{1}{3}}\sqrt{\tilde{d}}}{\sqrt{\tilde{a}}} \int  ^{\infty} _{u_{m}}\frac{{\rm d}u}{\sqrt {\frac{u_{m}^2} {   r_{th}^{4/3 }              }  (\ln{u_m}-\frac{3}{4})    \frac{u^{\frac{2}{3}}}{(\frac{3}{4}-\ln{u})^2}                                   +\frac{u^{\frac{8}{3}}}{(\frac{3}{4}-\ln{u}    )}       }}.
\end{equation}
To look at the physical content, we proceed to analyze the deflection angle  
  numerically.   Figure \ref{GL} depicts the dependence of the deflection angle on the closest distant of approach to the  wormhole for various lens parameters. 
 We find  $u_{m}=1.284$ which is a solution of the constraint (\ref{43}) by choosing $\frac{E^2}{L^2}=1$. We can see that the deflection angle $\theta$ increases as the parameter $r_{m}$ decreases.
\begin{center}
\begin{figure}[H] \hspace{4.cm}\includegraphics[width=8.cm]{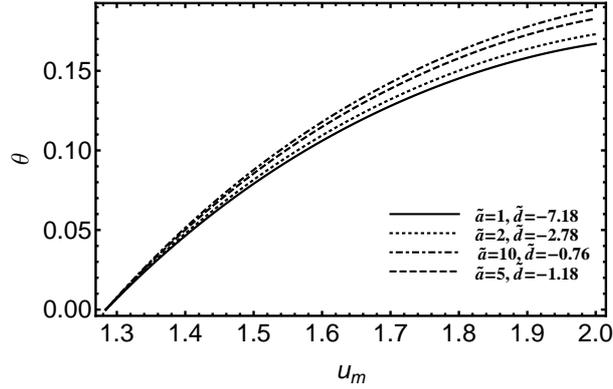}\caption{\label{GL} \small
Deflection angle $\theta$ as a function of  $u_{m}=\frac{r_{th}}{r_{m}}$ due to the magnetized wormhole for different lens parameters $\tilde{a}$ and $\tilde{d}$.}
\end{figure}
\end{center}

\section{ Magnetic Flux Along  the   Wormhole}\label{5}
In this section,  we calculate the magnetic flux across the two-dimensional  hypersurface $t$, $z=constant$. In general, the magnetic flux is given by the following Gaussian flux integral
\begin{equation}\label{36}
\Phi_{B}=\int F= \int^{r} _{r_{min}} \frac{1}{2} {F}^{\mu \nu}{\rm d}s_{\mu \nu},
\end{equation}
where $F^{\mu \nu}$ is the electromagnetic field tensor and ${\rm d}s_{\mu \nu}$ is  an element of two-dimensional surface area normal to the $z$-direction. Plugging  the only component of the electromagnetic tensor $F_{r \theta}$,  the two-surface element and evaluating the magnetic flux integral from the location of the throat  to a distance $r$, the
following magnetic flux will  result
\begin{equation}\label{37}
\Phi_{B}\left(r\right)=\frac{ 2 \pi \sqrt{\frac{\tilde{a}}{c}} }{\kappa}\{    \frac{1}{1+\tilde{d}\ln{r} }   -     \frac{4}{3\tilde{d} }  \}.
\end{equation}
Eq. (\ref{37}) is plotted in Figure \ref{fx11}
 for  $W=2$ which is defined as $W\equiv\frac{2 \pi \sqrt{\frac{\tilde{a}}{c}}}{\kappa }$ for the region $r_{throat }$ to a distance $r$. It should be  noted that  magnetic flux on  the other  side of  the magnetized cylindrical wormhole is highly concentrated around   $r=e^{-\frac{1}{\tilde{d}}}$. The total magnetic flux on one side of the wormhole from $r=r_{th}$ to $r=\infty$ is $\Phi_{total}=-\frac{8\pi \sqrt{\frac{\tilde{a}}{c}}}{3\kappa \tilde{d} },$

\begin{center}
\begin{figure}[H] \hspace{4.cm}\includegraphics[width=8.cm]{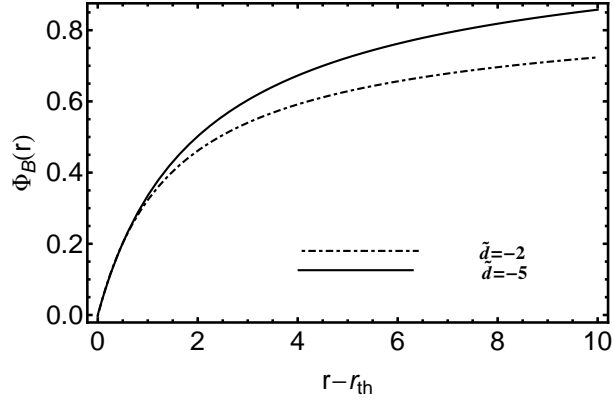}\caption{\label{fx11} \small
The magnetic flux on one side of  the magnetized cylindrical wormhole from the throat of the wormhole $r_{th}$ to $r$ with $W=2$ for two different values of the constant $\tilde{d}$.}
\end{figure}
\end{center}
\section{Energy Conditions}\label{6}
The $(3+1)$ dimensional energy-momentum tensor for our solution is diagonal  in the following form
  \begin{equation}\label{eq39}
  T^{\mu}_{~\nu}=diag(-\rho, P_r, P_\theta, P_z).
  \end{equation}
  The null energy condition\textbf{ (NEC) } is the assertion that for any null vector $k^\mu$, we have
  \begin{equation}
  T_{\mu \nu}k^\mu k^\nu \geq 0.
  \end{equation}
For the case of the energy-momentum tensor  (\ref{eq39}) this reduces to
 \begin{equation}\label{41}
 \rho+P_i \geq 0.
 \end{equation}
  The weak energy condition  (\textbf{WEC})
  asserts that for any timelike vector $v^\mu$
    \begin{equation}
  T_{\mu \nu}v^\mu v^\nu \geq 0.
  \end{equation}
  This condition can be written in the form
   \begin{equation}
 \rho \geq 0 \quad and  \quad \rho+P_i \geq 0.
 \end{equation}
  Finally the strong energy condition  (\textbf{SEC}) for any timelike vector $v^\mu$ is
   \begin{equation}
 ( T_{\mu \nu}-\frac{T}{2}g_{\mu \nu})v^\mu v^\nu \geq 0,
  \end{equation}
  in which $T$ is the trace of the energy-momentum tensor  \cite{14}. This condition can be stated in the following form by using the energy-momentum  tensor (\ref{eq39})
  \begin{equation}
 \rho+P_i \geq 0 \quad and  \quad \rho+\sum\limits_{i} P_i \geq 0.
 \end{equation}
  The Einstein field equations in terms  of the energy-momentum tensor (\ref{eq39}) for the metric (\ref{eq12})  lead to  the following expressions (c=G=1)
  \begin{equation}
  \rho=\frac{1}{8 \pi }\left[\frac{2}{9r^2}+\frac{\tilde{d}}{6r^2 (1+\tilde{d}\ln{r}) }  -\frac{3\tilde{d}^2}{4r^2  (1+\tilde{d}\ln{r})^2}\right],
  \end{equation}
  \begin{equation}
  P_r=\frac{1}{8 \pi }\left[\frac{\tilde{d}}{6r^2(1+\tilde{d}\ln{r})}-\frac{2}{9r^2}\right],
  \end{equation}
   \begin{equation}
  P_\theta=\frac{1}{8 \pi }\frac{4}{9r^2},
  \end{equation}
  and
   \begin{equation}
  P_z=\frac{1}{8 \pi }\frac{3\tilde{d}^2}{4r^2(1+\tilde{d}\ln{r})^2}.
  \end{equation}
Since the null energy condition is violated by the  matter that supports the static wormhole which is  \lq\lq{}exotic matter\rq\rq{}, here we only investigate the \textbf {NEC} for the wormhole solution \cite{-16}. Eq. (\ref{41}) yields
\begin{equation}
\rho+P_r=\frac{1}{8\pi}\left[\frac{1}{3r^2(\alpha+\ln{r})}-\frac{3}{4r^2(\alpha+\ln{r})^2}\right],
 \end{equation}
 \begin{equation}
\rho+P_{\theta}=\frac{1}{8\pi}\left[\frac{2}{3r^2}+\frac{1}{6r^2(\alpha+\ln{r})}-\frac{3}{4r^2(\alpha+\ln{r})^2}\right],
 \end{equation}
 and
  \begin{equation}
\rho+P_z=\frac{1}{8\pi}\left[\frac{2}{9r^2}+\frac{1}{6r^2(\alpha+\ln{r})}\right],
\end{equation}
in which  $\alpha \equiv \frac{1}{\tilde{d}}$. The contour  of the functions $\rho+P_r$, $\rho+P_{\theta}$, $\rho+P_z$ and $\rho$ with respect to the radial coordinate $r$ and constant $\alpha$ are depicted in Figures \ref{con3}-\ref{con1}.
 These figures show regions around the wormhole where the null energy condition is violated.
\begin{center}\begin{figure}[H]
\includegraphics[width=0.5\textwidth,height=0.3\textheight]{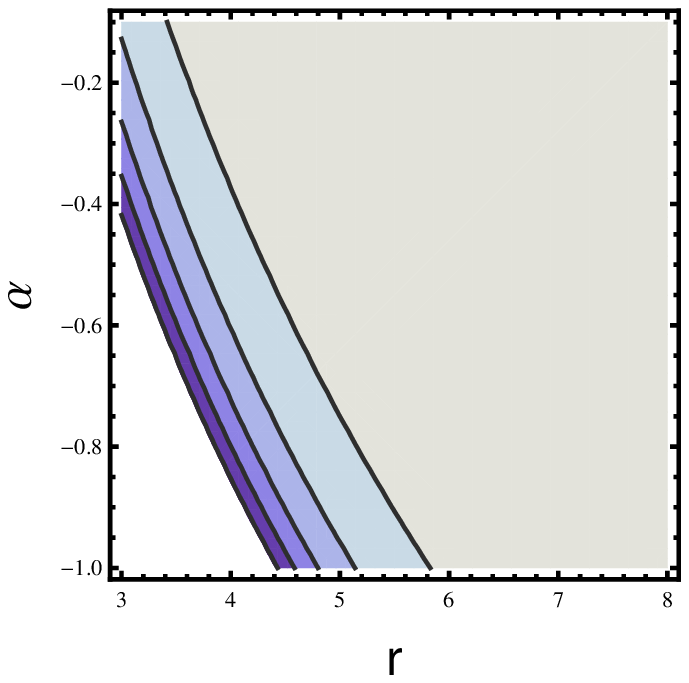}
\hspace{1cm}
\includegraphics[trim=-2cm -29 -29  2cm]{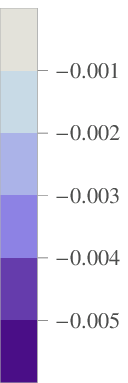}
\caption{ \small
The contour plot depicts the behavior of $\rho+P_r$ versus the radial coordinate $r$ and the parameter  $\alpha$. }\label{con1}
\end{figure}
\end{center}
\begin{center}\begin{figure}[H]
\includegraphics[width=0.5\textwidth,height=0.3\textheight]{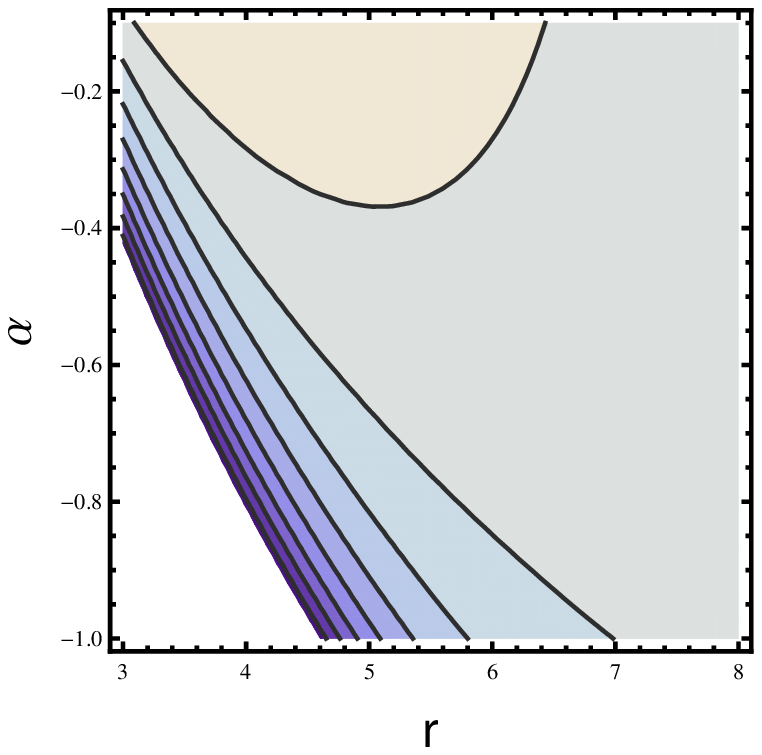}\label{con2}
\hspace{1cm}
\includegraphics
[trim=-2cm -23 -23  0cm]
{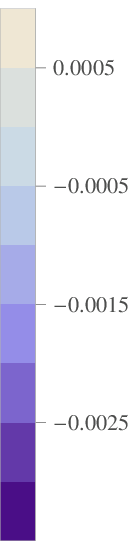}
\caption{ \small
The contour plot depicts the behavior of $\rho+P_{\theta}$ versus the radial coordinate $r$ and the parameter  $\alpha$. }
\end{figure}
\end{center}
\begin{center}\begin{figure}[H]
\includegraphics[width=0.5\textwidth,height=0.3\textheight]{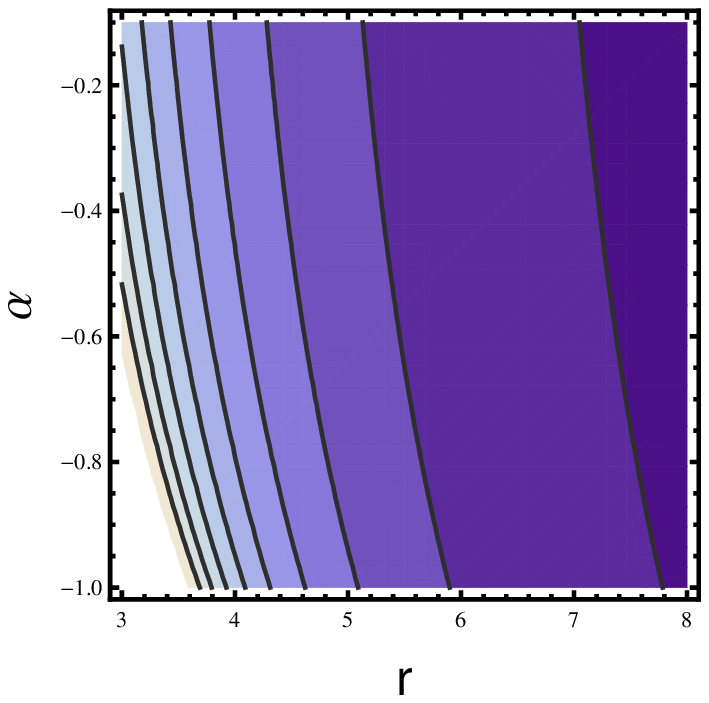}\label{con3}
\hspace{1cm}
\includegraphics
[trim=-2cm -29 -29  0cm]
{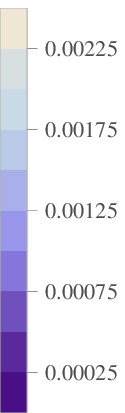}
\caption{ \small
The contour plot depicts the behavior of $\rho+P_{z}$ versus the radial coordinate $r$ and the parameter  $\alpha$. }
\end{figure}
\end{center}
\begin{center}\begin{figure}[H]
\includegraphics[width=0.5\textwidth,height=0.3\textheight]{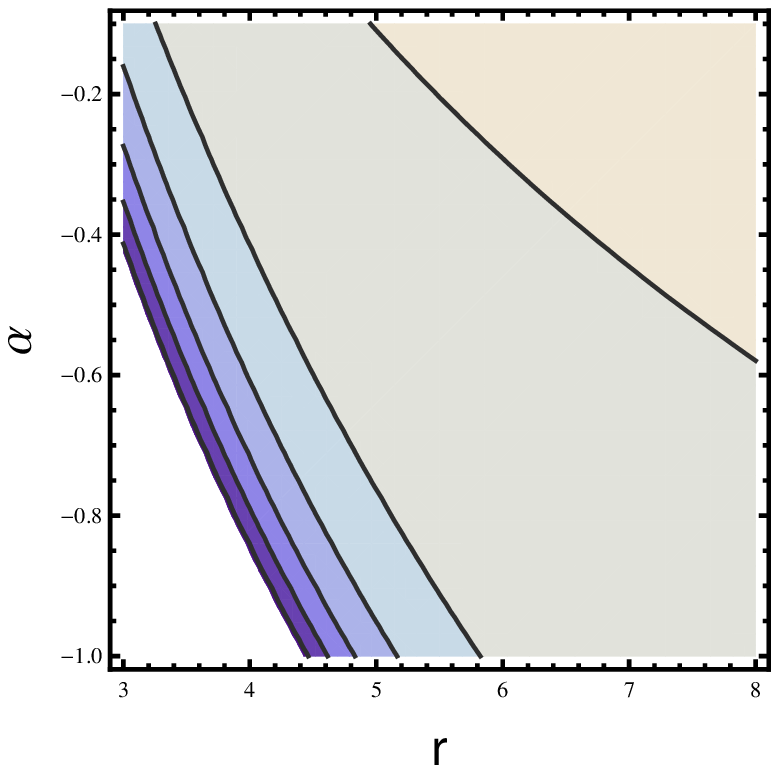}
\hspace{1cm}
\includegraphics
[trim=-2cm -29 -29  0cm]
{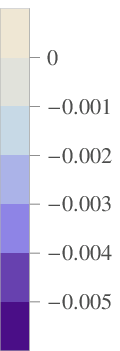}\caption{ \small
The contour plot depicts the behavior of $\rho$ versus the radial coordinate $r$ and the parameter  $\alpha$. }
\end{figure}
\end{center}

\section{Conclusion}
Starting from a cylindrical five-dimensional metric and using the standard Kaluza-Klein formalism, we derived an exact Ricci flat solution in five dimensions. We showed that in $3+1$ dimensions, this solution can be interpreted as a static, cylindrical wormhole supported by a scalar field and a magnetic field oriented along the magnetic wormhole (z-axis). We obtained the location of the throat, via minimizing the proper circumference of concentric circles around the wormhole. The magnetic flux on either side of the throat was calculated and shown to converge to a finite value on one side. In order to monitor the gravitational and electromagnetic effects of the wormhole, we built the equations of motion for a neutral and a charged test particle through the standard Lagrangian formalism. The repulsive character of the wormhole gravitational field was verified, by plotting the effective radial potential for the sample values of the particle energy and angular momentum. To study the gravitational lensing which is one of the most determinant  observational tests for confirming the existence of  wormholes we explicitly derived the photon orbit and the deflection angle as a function of the closest approach to the light rays.  By presenting  analytic expressions for the null energy condition, we show that this condition is violated.


\begin{thebibliography}{99}
\bibitem{0}
T. Kaluza, 966 (1921),
O. Klein, Z. Phys. 37, 895 (1926).
\bibitem{C. M. Chen}
C. M. Chen, arXiv preprint gr-qc/0009042.
\bibitem{-1}
P. Dobiasch and D. Maison, Stationary, Gen. Rel. Grav. 231-242, 14 (1982).
\bibitem{-2}
D. J. Gross and M. J. Perry,  Nucl. Phys. B226,  29-48 (1983).
\bibitem{riazi}
N. Riazi and S. Hashemi, arXiv preprint gr-qc/1405.4785.
\bibitem{1}
A. Einstein and N. Rosen, Phys. Rev. 48, 73 (1935).
\bibitem{-3}
J. A. Wheeler, Geons, Phys. Rev. 97, 511-536, (1955).
\bibitem{2}
F. S. Lobo, (2007),  arXiv preprint  gr-qc/ 0710.4474.
\bibitem{3}
M. S. Morris, K. S. Thorne, and U. Yurtsever, Phys. Rev. Lett. 61, 1446 (1988), M. S. Morris and K. S. Thorne, Am. J.
Phys. 56, 395 (1988).
\bibitem{-10}
P. K. Kuhfittig, Physical Review D, 71(10), 104007 (2005). 
\bibitem{-9}
J. P. de Leon, Journal of Cosmology and Astroparticle Physics, 11 (2009).
\bibitem{4}
S. W. Hawking,  W. Israel (Eds.),  Cambridge University Press, Cambridge, (1979).
\bibitem{5}
 M. Visser, Lorentzian Wormholes: From Einstein to Hawking (AIP, Woodbury, USA, 1995).
\bibitem{-111}S.V. Sushkov and S.-W. Kim,  Class. Quant. Grav. 19, 4909 (2002).

\bibitem{-12}
K. K. Nandi, B. Bhattacharjee, S. M. K. Alam and J. Evans,  Phys. Rev. D 57, 823 (1998).
\bibitem{-13}
L. A. Anchordoqui, A. G. Grunfeld and D. F. Torres,  Grav.Cosmol. 4, 287
(1998), [arXiv: gr-qc/9707025].
\bibitem{-14}
D. Hochberg,  Phys. Lett. 251, 349 (1990).
\bibitem{-15}
R. Garattini and F. S. N. Lobo, Class. Quant.
Grav. 24, 2401 (2007), [arXiv: gr-qc/0701020].
\bibitem{6}
S. B. Giddings and A. Strominger, Phys. Lett. B 230, 46 (1989); E. I. Guendelman, Gen. Rel. Grav. 23, 1415 (1991),
U. Bleyer, V. D. Ivashchuk, V. N. Melnikov, and A. Zhuk, Nucl. Phys. B 429, 177 (1994) [gr-qc/9405020], K. A. Bronnikov,
Grav. Cosmol. 2, 221 (1996) [gr-qc/9703020], K. A. Bronnikov, Grav. Cosmol. 4, 49 (1998) [hep-th/9710207], G. Dotti,
J. Oliva, and R. Troncoso, Phys. Rev. D 75, 024002 (2007) [hep-th/0607062], F. Canfora and A. Giacomini, Phys. Rev. D
78, 084034 (2008) [arXiv: 0808.1597 [hep-th]].
\bibitem{-4}
A. Chodos and S. Detweiler, Gen. Rel. Grav. 14, 879-890
(1982).
\bibitem{-5}
G. Clement,  Gen. Rel. Grav.
16,  477-489 (1984).
\bibitem{00}
S.I. Vacaru and D. Singleton, J. Math. Phys. 43 (2002).
\bibitem{01}
S.I. Vacaru, D. Singleton, V. A. Botan and D. A. Dotenco,  Phys. Lett. B 519
(2001).
\bibitem{02}
A.V. Aminova and P.I. Chumarov, Phys.
Rev. D 88 (2013).
\bibitem{12}
T. Kaluza, Sitz. Preuss. Akad. Wiss. K1, 966 (1921), O. Klein, Z. Phys. 37, 895
(1926),  T. Appelquist and A. Chodos, Phys. Rev. D28, 722 (1983), J.
Scherk and J.H. Schwarz, Nucl. Phys. B153, 61 (1979).
\bibitem{13}
M. Cvetic,  D. Youm, Nuclear Physics B, 438(1), 182-210 (1995). 
\bibitem{03}
A. Zee. Einstein Gravity in a Nutshell. (2013).
\bibitem{-17}
M. Visser and D. Hochberg. Annals of the Israel Physical Society, 13(gr-qc/9710001), 249-295,  (1997).
\bibitem{-11}
A. R. Prasanna,  and R. K. Varma.  Pramana-Journal of Physics 8.3, 229-244  (1977).
\bibitem{120}
K. Nakajima, H. Asada.  Physical Review D.  3, 85(10):107501, (2012).
\bibitem{121}
V. Perlick, Living Rev. Relativity 7, 9 (2004), http://relativity.livingreviews.org/Articles/lrr-2004-9.
\bibitem{122}
V. Perlick, arXiv preprint 1010.3416.
\bibitem{123}
P. Schneider, J. Ehlers and E. E. Falco, Gravitational Lenses (Springer-Verlag, Berlin, 1992).
\bibitem{124}
T. Harko, Z. Kovacs and F. S. N. Lobo, Phys. Rev. D 79, 064001 (2009).
\bibitem{125}
A. A. Abdujabbarov and B. J. Ahmedov, Astrophys. Space Sci. 321, 225, (2009).
\bibitem{126}
A. Pozanenko and A. Shatskiy, arXiv preprint 1007.3620.
\bibitem{127}
N. S. Kardashev, I. D. Novikov and A. A. Shatskiy, Int. J. Mod. Phys. D 16, 909 (2007).
\bibitem{128}
J. G. Cramer, R. L. Forward, M. S. Morris, M. Visser, G. Benford and G. A. Landis, Phys.Rev.
D 51, 3117 (1995).

\bibitem{129}
S. W. Kim and Y. M. Cho, in Evolution of the Universe and its Observational Quest (Universal
Academy Press, Tokyo, 1994), p. 353
\bibitem{130}
K. K. Nandi, Y. Z. Zhang and A. V. Zakharov, Phys. Rev. D 74, 024020 (2006).
\bibitem{131}
M. Safonova, D. F. Torres and G. E. Romero, Mod. Phys. Lett. A 16, 153 (2001).
\bibitem{132}
E. Eiroa, G. E. Romero and D. F. Torres, Mod. Phys. Lett. A 16, 973 (2001).
\bibitem{133}
M. Safonova, D. F. Torres and G. E. Romero, Phys. Rev. D 65, 023001 (2001).
\bibitem{134}
M. Safonova and D. F. Torres, Mod. Phys. Lett. A 17, 1685 (2002).
\bibitem{135}
F. Rahaman, M. Kalam and S. Chakraborty, Chin. J. Phys. 45, 518 (2007).
\bibitem{136}
A. A. Shatsuki, Astr. Rep. 48, 525, (2004).
\bibitem{137}
T. K. Dey and S. Sen, Mod. Phys. Lett. A 23, 953, (2008). 
\bibitem{14}
M. Visser, Lorentzian Wormholes: From Einstein to
Hawking, Springer, New York U.S. (1995).
\bibitem{-16}
M. K. Zangeneh, F. S.  Lobo, N.  Riazi,  Physical Review D, 90(2), 024072 (2014),

\end{thebibliography}
\end{document}